\documentclass[aps,pra,twocolumn,superscriptaddress,showpacs]{revtex4}
\usepackage{graphicx}

\newcommand{\mib}[1]{\mbox{\boldmath $#1$}}

\def\C{{\bf C}}
\def\Z{{\bf Z}}
\def\vq{{\bf q}}
\def\vp{{\bf p}}

\def\ve{{\bf e}}
\def\vsigma{\mib{\sigma}}

\begin{document}

\title{Quantum walks and 
orbital states of a Weyl particle}

\author{Makoto Katori} 
\email[]{katori@phys.chuo-u.ac.jp}
\affiliation{Department of Physics,
Faculty of Science and Engineering,
Chuo University, Kasuga, Bunkyo-ku, Tokyo 112-8551, Japan}
\author{Soichi Fujino}
\email[]{fujino@phys.chuo-u.ac.jp}
\affiliation{Department of Physics,
Faculty of Science and Engineering,
Chuo University, Kasuga, Bunkyo-ku, Tokyo 112-8551, Japan}
\author{Norio Konno}
\email[]{norio@mathlab.sci.ynu.ac.jp}
\affiliation{
Department of Applied Mathematics, 
Yokohama National University, \\
79-5 Tokiwadai, Yokohama 240-8501, Japan}

\date{May 18, 2005}

\begin{abstract}
The time-evolution equation of a one-dimensional quantum
walker is exactly mapped to the three-dimensional
Weyl equation for a zero-mass particle with spin 1/2,
in which each wave number $k$ of walker's wave function
is mapped to a point $\vq(k)$ in the three-dimensional 
momentum space and $\vq(k)$ makes a planar orbit 
as $k$ changes its value in $[-\pi, \pi)$.
The integration over $k$ providing the real-space
wave function for a quantum walker corresponds to considering
an orbital state of a Weyl particle, which is defined as a superposition 
(curvilinear integration) of the energy-momentum
eigenstates of a free Weyl equation along the orbit.
Konno's novel distribution function of
quantum-walker's pseudo-velocities in the long-time limit
is fully controlled by the shape of the orbit
and how the orbit is embedded in the 
three-dimensional momentum space.
The family of orbital states can be regarded as a geometrical
representation of the unitary group ${\rm U}(2)$
and the present study will propose a new group-theoretical
point of view for quantum-walk problems.

\end{abstract}


\maketitle

\section{INTRODUCTION} 

Quantum random walk models
\cite{ADZ93,Mey96,NV00,ABNVW01} have been intensively
studied as fundamental models in the basic theory of physics
to discuss the relationship between 
deterministic time-evolution with the probability interpretation
in quantum mechanics
and stochastic processes in statistical mechanics,
and in the quantum information theory to invent new algorithms 
for quantum computers
(see, \cite{Kem03,TFMK03,Amb03}, for recent reviews).
The interest in novel properties of quantum walks is now
spreading to many research fields and 
new applications are being reported. For example,
in probability theory a new type of limit theorems
quite different from the usual Gaussian-type central 
limit theorems was proved \cite{Konno02a,Konno02b,GJS04},
and in the solid-state physics of strongly correlated
electron systems the Landau-Zener transition dynamics
was related to a quantum walk \cite{OKAA04}.
This paper presents a new aspect of quantum walks, 
by showing that at each time $t \in [0, \infty)$
the quantum state of a one-dimensional quantum walker is
exactly mapped to that of a Weyl particle
(zero-mass particle with spin 1/2), which is
obtained by a curvilinear integration of 
energy-momentum eigenstates of a free Weyl equation
along a planar orbit appropriately embedded in the 
three-dimensional momentum space.
In a plane polar coordinate on an orbital plane,
$(q, \gamma),
q \in [0, \infty), \gamma \in [-\pi, \pi)$,
the equation of orbit is given by
\begin{equation}
\tan q= \frac{\sqrt{1-|a|^2}}{|a|} \frac{1}{\cos \gamma},
\label{eqn:orbit0}
\end{equation}
where $a \in \C$ with $|a| \in (0,1)$ is one of the parameters
of the unitary matrix in ${\rm U}(2)$,
which specifies the time-evolution of quantum walker.

In order to show the fact that the quantum random walk model
can be naturally considered to be 
a quantum-mechanical generalization
of usual random walk models,
and also in order to clarify key points of quantum-walk problems,
we first formulate the one-dimensional simple symmetric
random walk as a special case of the following
classical {\it random-turn} model.
(Such a model is also called the
correlated random walk \cite{Konno03}.)
Consider a walker at the origin of a one-dimensional
lattice $\Z=\{\cdots, -2,-1,0,1,2, \cdots\}$,
who is directed to the left with probability $q$
and to the right with probability $1-q$.
The walker has a coin, which is randomly tossed giving
heads with probability $p$ and tails with probability $1-p$.
He tosses the coin and, if the outcome is head,
he changes his direction, from the left to the right
or from the right to the left, and if the outcome is tail,
he does not change his direction.
In any case, then he makes a forward step.
At the new position, he tosses the coin again,
does or does not change his direction following the outcome
of the coin, and then make a forward step.
We assume that the walker repeats such random turns and steps 
$n$ times and the probability that he arrives at
the position $x \in \Z$ and also that he is directed to the left
(resp. right) after the $n$-th step is denoted by
$P_{n}^{(1)}(x)$ (resp. $P_{n}^{(2)}(x)$).
A simple application of the Fourier analysis 
gives
\begin{equation}
\left( \matrix{P_{n}^{(1)}(x) \cr P_{n}^{(2)}(x) } \right)
= \int_{-\pi}^{\pi} \frac{dk}{2 \pi}
e^{i k x} W(k)^{n} 
\left( \matrix{ q \cr 1-q } \right)
\label{eqn:classical1}
\end{equation}
with
\begin{equation}
W(k) =
\left( \matrix{e^{ik} & 0 \cr 0 & e^{-ik}} \right)
\left( \matrix{1-p & p \cr p & 1-p} \right),
\label{eqn:classical2}
\end{equation}
where $i=\sqrt{-1}$.
If the coin is fair ($p=1/2$),
the eigenvalues of the transition matrix
$W(k)$ are $\lambda=0$ and $\lambda=\cos k$, and
$P_{n}^{(1)}(x)=P_{n}^{(2)}(x) \equiv P_{n}(x)/2$
is independent of $q$ ({\it i.e.} independent
of the initial direction of walker).
This symmetric case realizes the simple symmetric random
walks, since 
$P_{n}(x)=\int_{-\pi}^{\pi} (dk/2\pi) e^{ikx} \cos^n k
={n \choose (n+x)/2}/2^n$.
Let temporal and spatial units be $\tau$ and $a$,
respectively, and set $n=t/\tau, x \to x/a, k \to a k$.
Then we consider the asymptotic of the probability density
$p_{t}(x)=P_{t/\tau}(x/a)/a$ in the continuum limit
$a, \tau \to 0$.
In the so-called diffusion scaling limit
$\tau=a^2 \to 0$, for
$(\cos a k)^{t/\tau}=(1-a^2 k^2/2+\cdots)^{t/\tau}
\to e^{-tk^2/2}$,
we find convergence of the probability density to
$$
p_{t}(x) = \int_{-\infty}^{\infty} \frac{dk}{2 \pi}
\exp \left[ - \frac{t}{2} k^2+i k x \right]
= \frac{1}{\sqrt{2 \pi t}} e^{-x^2/2t},
$$
which is called the heat-kernel, since it solves
the heat equation $[\partial/\partial t-
(1/2) \partial^2/\partial x^2] p_{t}(x)=0$
with the initial condition
$\lim_{t \to 0} p_{t}(x)=\delta(x)$.
The limit of walker's  position
is in the Gaussian distribution with 
mean zero and variance $t$,
and this convergence property of random-walk distribution
is a typical example of general central limit theorems.

Now we introduce the quantum random walk model as a 
quantum version of the above random-turn model.
In the LHS of (\ref{eqn:classical1}), 
the two component vector 
is replaced by a two-component wave function
\begin{equation}
  \Psi_{n}(x)=\left(
  \matrix{ \Psi_{n}^{(1)}(x) \cr
  \Psi_{n}^{(2)}(x) } \right),
\label{eqn:quantum1}
\end{equation}
and in the RHS of this equation, the initial distribution
of walker's directions 
$\left( \matrix{q \cr 1-q} \right)$ by an initial qubit
$\left( \matrix{\alpha \cr \beta} \right)$,
$\alpha, \beta \in \C, |\alpha|^2+|\beta|^2=1$,
the transition-probability matrix
$W(k)$ by a transition matrix of 
probability amplitudes
\begin{equation}
  U(k)=\left( \matrix{e^{ik} & 0 \cr 0 & e^{-ik}} \right) A.
\label{eqn:quantum2}
\end{equation}
Here $A$ is a $2 \times 2$ unitary matrix,
which determines behavior of a ``quantum coin".

In the context of quantum mechanics, the two-component
wave function (\ref{eqn:quantum1}) usually describes
quantum state of a particle with spin 1/2, 
and the Hilbert space of spin operators is spanned by the
Pauli matrices $\sigma_{j}, j=1,2,3$,
and the unit matrix $I_{2}$ of size 2.
In the classical random-turn model, the direction of step
is determined by the walker's direction that was randomly
changed just before doing the step.
Corresponding to this setting, operators which determine
motions of quantum walkers should be coupled with
spin operators acting ``spin states" of the walkers.
In the usual three-dimensional quantum mechanics,
the momentum vector operator
$\vp=(p_{1}, p_{2}, p_{3})=- i \hbar \nabla$
is the operator for the particle motions and the
easiest coupling with spin operators is given
in the form
\begin{equation}
{\cal H}(\vp)=\vsigma \cdot \vp,
\label{eqn:WeylH}
\end{equation}
where $\vsigma=(\sigma_{1}, \sigma_{2}, \sigma_{3})$.
This is known as the Hamiltonian for free motions
of a zero-mass particle with spin 1/2 like a neutrino.
The corresponding Schr\"odinger equation 
$$
i \hbar \frac{\partial}{\partial t} \hat{\Phi}_{t}(\vp)
=  (\vsigma \cdot \vp) \hat{\Phi}_{t}(\vp)
$$
is called the Weyl equation in the three-dimensional
momentum space \cite{Weyl29},
which is reduced from the Dirac equation for
four-component wave functions by setting the mass term to zero
(see, for example, Section 10.12 of \cite{BD64}
and Section 2-4-3 of \cite{IZ80}).

In the present paper, we will show that the time-evolution
equation of a quantum walker in the $k$-space
(one-dimensional Fourier space) is exactly mapped
to the Weyl equation in the $\vp$-space
(three-dimensional momentum space).
Since the quantum random walk model is defined on a lattice $\Z$,
the wave number $k$ should be in a finite region $[-\pi, \pi)$
with periodicity $k \pm 2\pi=k$ (the Brillouin zone).
In other words, the $k$-space can be identified with 
a unit circle, in which $k$ describes an angular coordinate
of a position on the circle.
We will show that, by our nonlinear mapping
$k \mapsto \vp(k)=(p_{1}(k), p_{2}(k), p_{3}(k))$,
this unit circle is transformed into a planar orbit
in the three-dimensional $\vp$-space.
The obtained orbit is no longer a unit circle, and the radial
coordinate $q$ depends on the direction on the
orbital plane. If we define the angular coordinate
$\gamma \in [-\pi, \pi)$ by an angle from the
direction to the ``perihelion" of orbit from the origin, 
the equation of orbit is given by (\ref{eqn:orbit0}).
The deformation of the unit circle, with which it
is embedded in the $\vp$-space, is fully described by
the Jacobian $J=|dk/d \gamma|$,
which appears when we consider the transformation of the integral
$\int_{-\pi}^{\pi}(dk/2\pi) f(k)$ to the curvilinear integral
along the orbit in the $\vp$-space.
One of the main results reported in the present paper is
that Konno's function \cite{Konno02a,Konno02b},
which describes the distribution of the pseudo-velocities
of quantum walker in the long-time limit $n \to \infty$,
is directly related with this Jacobian.
Moreover, it will be shown that, the normal direction
of the orbital plane depends on the choice of unitary
matrix $A$, and from this dependence the initial
qubit dependence is completely determined.

This paper is organized as follows.
In Sec.II, the precise description of the quantum random walk
models and the basic formulae for physical quantities
studied in this paper are given.
Konno's weak limit theorem is then briefly reviewed.
We rewrite $U(k)$ by using the exponential operators
with the Pauli matrices, and then 
the exact map to the Weyl equation is given in Sec.III.
There the one-to-one correspondence between 
the quantum-walker states in the $k$-space, 
each of which is specified by the unitary
matrix $A$, and the orbital states of a Weyl particle
in the three-dimensional $\vp$-space 
is clarified. Then in Sec.IV we follow
the argument by Grimmett {\it et al} \cite{GJS04}
and give a new proof of Konno's weak limit theorem.
Concluding remarks are given in Sec.V, and
Appendices are used for some details of calculations.

\section{THE MODEL AND KONNO'S DISTRIBUTION}

\subsection{Quantum random walk models associated with
unitary matrices}

Consider a two-component wave function (\ref{eqn:quantum1}).
Following the definition of the model given 
by Konno \cite{Konno02a,Konno02b}, we have
\begin{equation}
\Psi_{n+1}(x) 
= \left( \matrix{
a \Psi_{n}^{(1)}(x+1) + b \Psi_{n}^{(2)}(x+1) \cr
c \Psi_{n}^{(1)}(x-1) + d \Psi_{n}^{(2)}(x-1) } \right),
\label{eqn:U1}
\end{equation}
where
$A= \left( \matrix{a & b \cr
c & d } \right) \in {\rm U}(2)
\equiv$ a set of $2 \times 2$ unitary matrices.
If we set
$$
\hat{\Psi}_{n}(k) = \left( \matrix{
\hat{\Psi}_{n}^{(1)}(k) \cr \hat{\Psi}_{n}^{(2)}(k) } \right),
$$
and assume the relations
\begin{eqnarray}
\Psi_{n}^{(j)}(x) &=& \int_{-\pi}^{\pi} \frac{dk}{2 \pi}
e^{i k x} \hat{\Psi}_{n}^{(j)}(k), \nonumber\\
\hat{\Psi}_{n}^{(j)}(k) &=&
\sum_{x \in \Z} \Psi_{n}^{(j)}(x) e^{-i k x}, \quad
j=1,2,
\nonumber
\end{eqnarray}
then
(\ref{eqn:U1}) is rewritten 
in the wave-number space ($k$-space) of the 
Fourier transformation, $ k \in [-\pi, \pi)$ as
$$
\hat{\Psi}_{n+1}(k)=U(k) \hat{\Psi}_{n}(k), \quad n=0,1,2, \cdots,
$$
where $U(k)$
is given by (\ref{eqn:quantum2}).

The state at time-step $n$ is then obtained from the
initial state 
$\hat{\Psi}_{0}(k)=
\displaystyle{ \left( \matrix{ \alpha \cr \beta} \right) },
\alpha, \beta \in \C, |\alpha|^2+|\beta|^2=1,$ by
\begin{equation}
\hat{\Psi}_{n}(k) =U(k)^{n} \hat{\Psi}_{0}(k),
\label{eqn:evolution1}
\end{equation}
whose Fourier transformation gives the wave function at
time-step $n$ as
\begin{eqnarray}
\Psi_{n}(x)
&=& \int_{-\pi}^{\pi} \frac{dk}{2 \pi}
e^{ikx} \hat{\Psi}_{n}(k) \nonumber\\
&=& \int_{-\pi}^{\pi} \frac{dk}{2 \pi}
e^{ikx} U(k)^{n} \hat{\Psi}_{0}(k).
\nonumber
\end{eqnarray}
The distribution function in the real space at time-step $n$
is given by
\begin{eqnarray}
P_{n}(x) &=& |\Psi_{n}(x)|^2 
= \Psi_{n}^{\dagger}(x) \Psi_{n}(x) \nonumber\\
&=& \int_{-\pi}^{\pi} \frac{dk'}{2 \pi}
\int_{-\pi}^{\pi} \frac{dk}{2 \pi}
e^{i(k-k')x} \nonumber\\
&& \quad \times
\Big( \hat{\Psi}^{\dagger}_{0}(k') U^{\dagger}(k')^{n} \Big)
\Big(U(k)^{n} \hat{\Psi}_{0}(k)\Big).
\label{eqn:Pn}
\end{eqnarray}

It is known that any matrix in ${\rm U}(2)$ can be 
parameterized as a matrix in ${\rm SU}(2)$ times 
a phase factor in the form 
$e^{i \varphi}, \varphi \in [-\pi/2, \pi/2)$,
where ${\rm SU}(2)$ denotes the set of $2 \times 2$
unitary matrices with determinant 1.
For example, the Hadamard matrix 
\begin{equation}
H=\frac{1}{\sqrt{2}} \left( \matrix{1 & 1 \cr 1 & -1} \right)
\in {\rm U}(2)
\label{eqn:Hadamard}
\end{equation}
is written as
$
 H=e^{-i \pi/2} A
$
with
$
  A= \frac{1}{\sqrt{2}} \left( \matrix{
  i & i \cr i & -i} \right) \in {\rm SU}(2).
$
Since the phase factor $e^{i \varphi}$ 
of $U(k)$ is irrelevant in calculating the
distribution function (\ref{eqn:Pn}), without loss of
generality, we can assume that the matrix $A$ 
in (\ref{eqn:quantum2}) is chosen from ${\rm SU}(2)$.
We can see that ${\rm SU}(2)$ is generally
written as 
\begin{eqnarray}
{\rm SU}(2) &=& 
\left\{ A=\left( \matrix{a & b \cr -b^{*} & a^{*}} \right);
a, b \in \C, |a|^2+|b|^2=1 \right\} \nonumber\\
&=& \left\{
A=\left( \matrix{u e^{i \theta} & \sqrt{1-u^2} e^{i \phi} \cr
-\sqrt{1-u^2} e^{-i \phi} & u e^{-i \theta}} \right); 
\right. \nonumber\\
&& \qquad \qquad \quad 
u \in [0,1], \theta, \phi \in [-\pi, \pi) \Big\}.
\label{eqn:K1}
\end{eqnarray}
As shown by the second equality above,
it is a three-dimensional space 
parameterized by real variables
$u, \theta$ and $\phi$ (Cayley-Klein parameters).

\subsection{Formulae of expectations}

Let $X_{n}$ denote the position of the one-dimensional
quantum walk at time-step $n =0,1,2, \cdots$
and consider a function $f$ of $x \in \Z$.
The expectation of $f(X_{n})$ is defined by
\begin{eqnarray}
&& \Big\langle f(X_{n}) \Big\rangle
= \sum_{x \in \Z} f(x) P_{n}(x) \nonumber\\
&& \, = \sum_{x \in \Z} f(x) 
\int_{-\pi}^{\pi} \frac{dk'}{2 \pi} e^{-i k' x}
\hat{\Psi}_{n}^{\dagger}(k') 
\int_{-\pi}^{\pi} \frac{dk}{2\pi}
e^{i k x} \hat{\Psi}_{n}(k).
\nonumber
\end{eqnarray}
If 
$f(x)=x^r, r=0,1,2, \cdots$, it is written as 
\begin{eqnarray}
\Big\langle X_{n}^{r} \Big\rangle 
&=& \sum_{x \in \Z}
\int_{-\pi}^{\pi} \frac{dk'}{2 \pi} e^{-i k' x}
\hat{\Psi}_{n}^{\dagger}(k') \nonumber\\
&& \times
\int_{-\pi}^{\pi} \frac{dk}{2 \pi}
\left\{ \left(-i \frac{d}{dk} \right)^{r}
e^{i k x} \right\} \hat{\Psi}_{n}(k).
\nonumber
\end{eqnarray}
We note that
$\hat{\Psi}_{n}(k)$ should be a periodic function
of $k \in [-\pi, \pi)$, and then by partial integrations,
we will have
\begin{eqnarray}
&& \int_{-\pi}^{\pi} \frac{dk}{2\pi}
\left\{ \left(-i \frac{d}{dk} \right)^{r} 
e^{ikx} \right\} \hat{\Psi}_{n}(k) \nonumber\\
&& \qquad 
= \int_{-\pi}^{\pi} \frac{dk}{2 \pi} e^{ikx}
\left(i \frac{d}{dk} \right)^{r} \hat{\Psi}_{n}(k),
\nonumber
\end{eqnarray}
and thus
$$
\Big\langle X_{n}^{r} \Big\rangle
= \int_{-\pi}^{\pi} \frac{dk}{2 \pi}
\hat{\Psi}_{n}^{\dagger}(k) \left( i \frac{d}{dk} \right)^{r}
\hat{\Psi}_{n}(k),
$$
where we have used the summation formula
$\sum_{x \in \Z} e^{ix k}=2 \pi \delta(k)$.
Then, if $f(x)$ is analytic around $x=0$, that is,
if it has a converging Taylor expansion in the form
$f(x)=\sum_{j=0}^{\infty} a_{j} x^{j}$,
we will have the formula
\begin{equation}
\Big\langle f(X_{n}) \Big\rangle
= \int_{-\pi}^{\pi} \frac{dk}{2\pi}
\hat{\Psi}_{n}^{\dagger}(k) 
f \left( i \frac{d}{dk} \right)
\hat{\Psi}_{n}(k).
\label{eqn:expec2}
\end{equation}

\subsection{Konno's results}

A fundamental requirement of the quantum mechanics is
that any time-evolution of an isolated system is given
by a unitary transformation of wave function,
which is necessary to allow us to adopt the usual formula
for probability distribution of outcomes in observations
given by using squares of the wave function.
In the present case, $U(k)$ is unitary and 
the probability that the quantum walker is observed 
at the position $x \in \Z$ after $n$-th step
is given by (\ref{eqn:Pn}).
The expectation of a function $f$ of position 
at time-step $n$ is then calculated as (\ref{eqn:expec2}).

It should be noted here that the unitarity of $U(k)$
ensures $|\lambda|=1$ for any eigenvalue $\lambda$ of it.
This fact implies that in principle 
we are not able to find any
convergence property of wave function $\Psi_{n}(x)$
nor of the probability $P_{n}(x)$
in the long-time limit $n \to \infty$ 
in quantum-walk problems.
It presents a striking contrast to the classical systems of
random walks, which will generally converge to 
diffusion particle systems in the long-time and
large-scale limit, as we have demonstrated in Sec.I
in the simple and symmetric case.

Recently, however, Konno proved that the distribution
of the pseudo-velocities of the quantum walker $X_{n}/n$ 
{\it does} converge in such a weak sense 
that any moment of the pseudo-velocity
converges in the long-time limit $n \to \infty$ 
to a moment of a random variable, whose 
distributed is given by a novel probability 
density function \cite{Konno02a,Konno02b}.
That is, for any $r=0,1,2, \cdots$,
$$
\langle (X_{n}/n)^{r} \rangle
\, \to \, 
\int_{-\infty}^{\infty} dy \, y^{r} \nu(y) \quad
\mbox{in} \quad n \to \infty,
$$
where
\begin{equation}
\nu(y)=\mu(y; |a|) {\cal I}(y; a,b; \alpha, \beta) 
{\bf 1}_{\{|y| < |a|\}}
\label{eqn:Konno2}
\end{equation}
with
\begin{eqnarray}
\label{eqn:Konno3}
&& \mu(y; |a|) = \frac{\sqrt{1-|a|^2}}
{\pi (1-y^2) \sqrt{|a|^2-y^2}}, \\
&& {\cal I}(y; a, b; \alpha, \beta) \nonumber\\
\label{eqn:Konno4}
&& \,= 1- \left( 
|\alpha|^2-|\beta|^2 
+\frac{\alpha \beta^{*} a b^{*}+ \alpha^{*} \beta a^{*} b}
{|a|^2} \right) y.
\end{eqnarray}
Here ${\bf 1}_{\{\omega\}}$ denotes the indicator function
of a condition $\omega$;
${\bf 1}_{\{\omega\}}=1$ if $\omega$ is satisfied
and ${\bf 1}_{\{\omega\}}=0$, otherwise.
The special case, in which the matrix $A$ is 
(proportional to) the Hadamard matrix
(\ref{eqn:Hadamard}), 
has been studied well with the name of {\it Hadamard walk},
since it corresponds to the $p=1/2$ case of the
classical random-turn model (\ref{eqn:classical1}) with
(\ref{eqn:classical2}). 
But Konno's {\it weak limit theorem}
claims that even in this case, ${\cal I} \not=1$;
the limit distribution depends on the initial qubit
and it is not symmetric in general.
The limit distribution is very sensitive to the initial
qubit, and if and only if the initial qubit is chosen as
$|\alpha|=|\beta|$ and 
$\alpha \beta^{*} a b^{*}+\alpha^{*} \beta a^{*} b=0$
for given $a$ and $b$ in $A$,
probability density becomes a symmetric function
$\mu(y; |a|) {\bf 1}_{\{|y| < |a|\}}$
(see \cite{Konno02a,Konno02b}).

\section{MAP TO WEYL EQUATION}

\subsection{Exponential operator with Pauli matrices}

Consider the Pauli matrices,
$\sigma_{1}=\left( \matrix{0 & 1 \cr 1 & 0} \right)$,
$\sigma_{2}=\left( \matrix{0 & -i \cr i & 0} \right)$,
$\sigma_{3}=\left( \matrix{1 & 0 \cr 0 & -1} \right)$.
For an arbitrary three-dimensional vector 
$\vq=(q_{1}, q_{2}, q_{3})$, we can see that
$(\vsigma\cdot \vq)^2
=q^2 I_{2}$,
where $q=|\vq|$.
Then
\begin{eqnarray}
e^{-i \vsigma\cdot \vq}
&=& \sum_{n=0}^{\infty} \frac{1}{n!} 
(-i \vsigma\cdot \vq)^{n} \nonumber\\
&=& \Bigg( I_{2}-i \vsigma \cdot
\hat{\vq} \, \tan q \Bigg) \cos q,
\label{eqn:exp1}
\end{eqnarray}
where $\hat{\vq}$ is a unit vector defined by
$\hat{\vq}=\vq/q.$
\begin{widetext}
By using the well-known algebra of the Pauli matrices,
we find that (\ref{eqn:quantum2}) 
with $A \in {\rm SU}(2)$ can be
written as
\begin{eqnarray}
   U(k) 
   &=& \left( \matrix{
   u e^{i(k+\theta)} & \sqrt{1-u^2} e^{i(k+\phi)} \cr
   -\sqrt{1-u^2} e^{-i(k+\phi)} & u e^{-i(k+\theta)} } \right)
   \nonumber\\
&=& u \cos(k+\theta) \left[ I_{2} 
+ i \left( \frac{\sqrt{1-u^2}}{u}
\frac{\sin(k+\phi)}{\cos(k+\theta)} \sigma_{1}
+\frac{\sqrt{1-u^2}}{u}
\frac{\cos(k+\phi)}{\cos(k+\theta)} \sigma_{2}
+\tan (k+\theta) \, \sigma_{3} \right) \right].
\label{eqn:Uk3}
\end{eqnarray}
\end{widetext}
In order to identify $U(k)$ with (\ref{eqn:exp1})
by choosing the vector ${\bf q}$ appropriately,
we consider a system of equations;
$u \cos (k+\theta) = \cos q$,
$(\sqrt{1-u^2}/u) (\sin(k+\phi)/\cos(k+\theta))
=-\hat{q}_{1} \tan q$,
$(\sqrt{1-u^2}/u)
(\cos(k+\phi)/\cos(k+\theta))
=-\hat{q}_{2} \tan q$,
$\tan (k+\theta) = -\hat{q}_{3} \tan q$.
It is solved as
\begin{eqnarray}
\label{eqn:q01}
q(k) &=& \arccos \Bigg[ u \cos (k+\theta) \Bigg] \\
\label{eqn:q02}
&=&
\arctan \left[
\frac{1}{\cos(k+\theta)}
\sqrt{ \frac{1}{u^2}-\cos^2(k+\theta)} \right], 
\end{eqnarray}
where we take the principal value of $\arccos x$
and choose appropriate branches of $\arctan x$
so that $q(k)$ is a periodic and continuous function
of $k \in [-\pi, \pi)$,
and
\begin{eqnarray}
\hat{q}_{1}(k) &=&
- \frac{\sqrt{1-u^2}/u}
{\sqrt{1/u^2-\cos^2(k+\theta)}}
\sin(k+\phi), \nonumber\\
\hat{q}_{2}(k) &=& - \frac{\sqrt{1-u^2}/u}
{\sqrt{1/u^2-\cos^2(k+\theta)}}
\cos(k+\phi), \nonumber\\
\hat{q}_{3}(k) &=&
- \frac{1}
{\sqrt{1/u^2-\cos^2(k+\theta)}}
\sin(k+\theta).
\label{eqn:q03}
\end{eqnarray}
We then define a three-dimensional vector
\begin{equation}
\vq(k)=\Big(q(k) \hat{q}_{1}(k),
q(k)\hat{q}_{2}(k), q(k)\hat{q}_{3}(k) \Big).
\label{eqn:map}
\end{equation}
By the formula (\ref{eqn:exp1}), $U(k)$ is now
written as
\begin{equation}
  U(k) = e^{-i \vsigma\cdot \vq(k)}.
\label{eqn:Uk4}
\end{equation}

We can interpolate time-steps $n=0,1,2, \cdots$ 
to define a continuous time $t \in [0, \infty)$,
and we have
\begin{equation}
\hat{\Psi}_{t}(k) =
e^{-i t \vsigma \cdot \vq(k)}
\hat{\Psi}_{0}(k), \quad
k \in [-\pi, \pi),
\label{eqn:Psit}
\end{equation}
which solves the Weyl equation with $\hbar=1$, 
\begin{equation}
i \frac{\partial}{\partial t} 
\hat{\Psi}_{t}(k)=
{\cal H}(\vq(k)) \hat{\Psi}_{t}(k),
\label{eqn:WeylEq}
\end{equation}
where the Hamiltonian is given by (\ref{eqn:WeylH}).
The equations (\ref{eqn:q01})-(\ref{eqn:map}) 
define a map $k \in [-\pi, \pi) \mapsto
\vq \in$ $\vp$-space, and by this map
the time-evolution equation (\ref{eqn:evolution1})
of the one-dimensional quantum walk is
exactly transformed to 
the Weyl equation (\ref{eqn:WeylEq}).

It is easy to solve the eigenvalue problem of the
Hamiltonian (\ref{eqn:WeylH}) 
of a free Weyl particle with momentum 
$\vp$ \cite{BD64,IZ80}. The eigenvalues are
$\lambda= \pm p$, where $p=|\vp|$. 
The corresponding energy-momentum eigenfunctions are
independent of the absolute value of $\vp$ and
given as functions of $\hat{\vp}=\vp/p$ by
\begin{eqnarray}
\psi_{+}(\hat{\vp}) &=& \left( \matrix{
\psi_{+}^{(1)}(\hat{\vp}) \cr 
\psi_{+}^{(2)}(\hat{\vp}) } \right)
= \left( \matrix{ \cos (\theta_{p}/2) \cr
\sin (\theta_{p}/2) e^{i \varphi_{p}} } \right),
\nonumber\\
\psi_{-}(\hat{\vp}) &=& \left( \matrix{
\psi_{-}^{(1)}(\hat{\vp}) \cr 
\psi_{-}^{(2)}(\hat{\vp}) } \right)
= \left( \matrix{ 
-\sin (\theta_{p}/2) e^{-i \varphi_{p}} \cr
\cos (\theta_{p}/2) } \right)
\label{eqn:eigenf2}
\end{eqnarray}
in the polar coordinate of momentum,
$p_{1}=p \sin \theta_{p} \cos \varphi_{p}$, 
$p_{2}= p \sin \theta_{p} \sin \varphi_{p}$,
$p_{3} = p \cos \theta_{p}$,
which are orthonormal in the sense that
\begin{eqnarray}
&& |\psi_{+}(\hat{\vp})|^2 
= \psi_{+}^{\dagger}(\hat{\vp}) 
\psi_{+}(\hat{\vp}) =1,
\nonumber\\
&& |\psi_{-}(\hat{\vp})|^2 
= \psi_{-}^{\dagger}(\hat{\vp}) 
\psi_{-}(\hat{\vp}) =1,
\nonumber\\
&& \psi_{+}^{\dagger}(\hat{\vp}) \psi_{-}(\hat{\vp})
= \psi_{-}^{\dagger}(\hat{\vp}) \psi_{+}(\hat{\vp}) =0.
\label{eqn:eigenf3}
\end{eqnarray}
Now the problem is reduced to the study of the property of
nonlinear map $k \mapsto \vq \in$ $\vp$-space.

\subsection{Planar orbital states in the
momentum space}

Since (\ref{eqn:q01})-(\ref{eqn:q03}) show that all the components
$q_{j}(k), j=1,2,3$ are periodic functions of $k \in [-\pi,\pi)$,
$\vq(k)$ gives a one-parameter orbit in the three-dimensional
$\vp$-space parameterized by
$k \in [-\pi, \pi)$.
Define a unit vector
\begin{equation}
\hat{\ve}_{3}= \Big( -u \cos(\phi-\theta), u \sin(\phi-\theta),
\sqrt{1-u^2} \Big).
\label{eqn:n}
\end{equation}
Then we can see
$ \vq(k) \cdot \hat{\ve}_{3}=0$.
That is,
$$
   \vq(k) \quad \perp \quad \hat{\ve}_{3} \qquad
   \mbox{for all} \quad k \in [-\pi, \pi),
$$
which implies that the orbit is on a
plane including the origin, whose normal vector
is $\hat{\ve}_{3}$.
We denote this orbital plane by
$\Pi(u, \theta, \phi)$.

Since we have assumed the principal value for
$\arccos x$, $0 \leq \arccos u \leq \pi/2$
for any given $0 \leq u \leq 1$,
where $\arccos u=\pi/2$ if $u=0$,
and $\arccos u=0$ if $u=1$.
As explained in Appendix A, we can see that
$q(k)=|\vq(k)|$ takes its minimum value
$q_{\rm min}=\arccos u$ when $k=-\theta$
(mod $2 \pi$)
and its maximum value $q_{\rm max}=\pi-q_{\rm min}$
when $k=-\pi-\theta$ (mod $2 \pi$).
When $k=-\pi/2-\theta$ and $k=\pi/2-\theta$ (mod $2 \pi$),
$q(k)=\pi/2$.
Let
\begin{eqnarray}
\hat{\ve}_{1} &=&
\Big( - \sin(\phi-\theta),
-\cos(\phi-\theta), 0 \Big), \nonumber\\
\hat{\ve}_{2} &=& \Big( \sqrt{1-u^2} \cos (\phi-\theta),
- \sqrt{1-u^2} \sin(\phi-\theta), u \Big),
\nonumber\\
\label{eqn:ve}
\end{eqnarray}
so that a set $(\hat{\ve}_{1}, \hat{\ve}_{2}, \hat{\ve}_{3})$
makes an orthonormal basis in the $\vp$-space
and the orbital plane $\Pi(u, \theta, \phi)$ is equal to 
the $(\hat{\ve}_{1}, \hat{\ve}_{2})$-plane.
The unit vector $\hat{\ve}_{1}$ is directed
to the ``perihelion" $\vq(k_{0})$ with
$k_{0}=-\theta$ (mod $2 \pi$) and we 
define the angle $\gamma$ in this plane by
$$
\cos \gamma= \hat{\vq}(k) \cdot \hat{\ve}_{1},
$$
where $\hat{\vq}(k)=\vq(k)/q(k)$.
Then we have
\begin{eqnarray}
\label{eqn:cosg}
\cos \gamma &=& \frac{(\sqrt{1-u^2}/u) \cos (k+\theta)}
{\sqrt{1/u^2 - \cos^2 (k+\theta)}}, \\
\label{eqn:sing}
\sin \gamma &=& - \frac{(1/u) \sin (k+\theta)}
{\sqrt{1/u^2 - \cos^2(k+\theta)}}.
\end{eqnarray}
Combining (\ref{eqn:q02}) and (\ref{eqn:cosg}), we have
the equation for the orbit
\begin{equation}
\tan q= \frac{\sqrt{1-u^2}}{u}
\frac{1}{\cos \gamma}
\label{eqn:orbit}
\end{equation}
in the plane
polar coordinates $(q, \gamma),
0 \leq q < \infty, \gamma \in [-\pi, \pi)$, on the 
orbital plane $\Pi(u, \theta, \phi)$.
We can rewrite (\ref{eqn:q01})-(\ref{eqn:map})
as the functions of $q$ and $\gamma$ as
\begin{eqnarray}
q_{1} &=& q \sqrt{1-u^2} \cos(\phi-\theta) \sin \gamma
- q \sin (\phi-\theta) \cos \gamma, \nonumber\\
q_{2} &=& - q \sqrt{1-u^2} \sin (\phi-\theta) \sin \gamma 
-q \cos (\phi-\theta)\cos \gamma, \nonumber\\
q_{3} &=& q u \sin \gamma,
\label{eqn:q3qg}
\end{eqnarray}
which are the representations for
the three components of
$\vq$ by the plane polar coordinates
$(q, \gamma)$
on $\Pi(u, \theta, \phi)$.

\begin{figure}[htpb]
\includegraphics[width=0.6\linewidth]{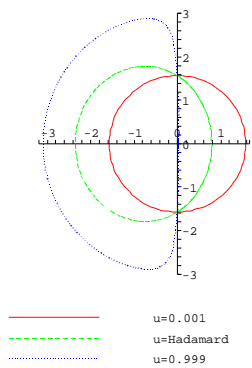}
\caption{(Color online)
The dependence on $u$ of shape of the orbit.
Three cases $u \simeq 0$, $u=1/\sqrt{2}$
(the Hadamard matrix), and $u \simeq 1$ are
shown.
\label{fig:qwFig}}
\end{figure}

It is interesting to see dependence of the orbit
$\{\vq(k); k \in [-\pi, \pi)\}$ on the
parameters $u, \theta, \phi$ of unitary matrices.
The normal direction of orbital plane
$\hat{\ve}_{3}$ given by (\ref{eqn:n}) is in the $z$-direction,
when $u=0$, and it is tilted, as $u$ increases from 0 to 1,
to the 
$(-\cos (\phi-\theta), \sin(\phi-\theta),0)$-direction 
lying in the $(x,y)$-plane.
We assume that $\arctan x$ takes a principal value
($0 \leq \arctan x < \pi/2$) for $x \geq 0$,
and a value in $( \pi/2, \pi)$ for $x < 0$.
Then (\ref{eqn:orbit}) gives
$$
q= \arctan \left[ \frac{\sqrt{1-u^2}}{u}
\frac{1}{\cos \gamma} \right].
$$
When $u=0$, $q \equiv \pi/2$, which implies that
the orbit is a circle with the radius $\pi/2$
and the center at the origin.
Since $\cos \gamma \geq 0$ (resp. $\cos \gamma < 0$)
for $\gamma \in [-\pi/2, \pi/2]$
(resp. $\gamma \in [-\pi, -\pi/2) \cup
(\pi/2, \pi)$),
in the limit $u \to 1$, we will see that
$q \to 0$ when $- \pi/2 \leq \gamma \leq \pi/2$
and $q \to \pi$ when $-\pi \leq \gamma < -\pi/2$
and $\pi/2 < \gamma < \pi$.
When $u \in (0, 1)$, the orbit is a deformed circle.
Therefore, if we observe the orbit 
riding on the tilting orbital plane,
its shape is changing in $u$ as shown in Fig. \ref{fig:qwFig}.

\subsection{Curvilinear integration along the orbit}

Now we consider the integration with respect to
$k$ on $[-\pi, \pi)$,
which is necessary to calculate the expectations 
(\ref{eqn:expec2}).
The above result implies that
this integration corresponds to the integration
over $\gamma \in [-\pi, \pi)$,
which performs the curvilinear integration 
along the orbit.
We write the Jacobian associated with the
map $k \to \gamma$ as $J=|dk/d\gamma|$.
By simple calculation, we have (see Appendix A),
\begin{equation}
  J= \frac{\sqrt{1-u^2}}{1-u^2 \sin^2 \gamma},
\label{eqn:J}
\end{equation}
and the relation
\begin{equation}
  \int_{-\pi}^{\pi} \frac{dk}{2 \pi} f(k)
= \int_{-\pi}^{\pi} \frac{d \gamma}{2 \pi}
\frac{\sqrt{1-u^2}}{1-u^2 \sin^2 \gamma} f(k(\gamma))
\label{eqn:dkdg}
\end{equation}
is established, where $k(\gamma)$ is determined
by (\ref{eqn:sinkt}) and (\ref{eqn:coskt}).

Then we change the variable as
\begin{equation}
\gamma \to y \quad; \quad
y=u \sin \gamma,
\label{eqn:defy}
\end{equation}
to have the equality
$$
  \int_{-\pi}^{\pi} \frac{dk}{2 \pi} f(k)
  = 2 \int_{-u}^{u} \frac{dy}{2 \pi}
  \frac{1}{\sqrt{u^2-y^2}}
  \frac{\sqrt{1-u^2}}{1-y^2} f(k(y)).
$$
That is, we arrive at the formula
$$
\int_{-\pi}^{\pi} \frac{dk}{2 \pi} f(k)
= \int_{-u}^{u} dy \, \mu(y; u) f(k(y))
$$
with Konno's function (\ref{eqn:Konno3}).

\section{LIMIT THEOREM OF QUANTUM WALKS}

\subsection{Quantum walk starting from the origin}

We assume that the quantum walk starts from $x=0$, that is 
$$
\Psi_{0}(x)=\delta(x) 
\left( \matrix{ \alpha \cr \beta } \right)
\quad \Leftrightarrow \quad
\hat{\Psi}_{0}(k)= \left( \matrix{ \alpha \cr \beta } \right),
$$
where $\alpha, \beta \in \C$ with
$|\alpha|^2+|\beta|^2=1$.
We can express this initial wave function $\hat{\Psi}_{0}(k)$
as a linear combination of the two eigenfunctions
(\ref{eqn:eigenf2})
of the Hamiltonian ${\cal H}({\bf p})$ with
${\bf p}=\vq(k)$ in the way
\begin{equation}
\hat{\Psi}_{0}(k)=C_{+}(\hat{\vq}(k)) \psi_{+}(\hat{\vq}(k))
+ C_{-}(\hat{\vq}(k)) \psi_{-}(\hat{\vq}(k)),
\label{eqn:expansion1}
\end{equation}
where
$C_{+}(\hat{\vp})$ and $C_{-}(\hat{\vp})$ are
given as functions of $\hat{\vp}=\vp/|\vp|$ 
by (\ref{eqn:Cpmp1}) in Appendix B.
Then, for $n=0,1,2, \cdots$, (\ref{eqn:Psit}) gives
\begin{eqnarray}
\hat{\Psi}_{n}(k) 
&=& e^{-i {\cal H}(\vq(k))n} \nonumber\\
&& \times
\Bigg\{C_{+}(\hat{\vq}(k)) \psi_{+}(\hat{\vq}(k))
+ C_{-}(\hat{\vq}(k)) \psi_{-}(\hat{\vq}(k)) \Bigg\} \nonumber\\
&=&
e^{-i q(k) n} C_{+}(\hat{\vq}(k)) \psi_{+}(\hat{\vq}(k))
\nonumber\\
&& \qquad \qquad 
+ e^{i q(k) n} C_{-}(\hat{\vq}(k)) \psi_{-}(\hat{\vq}(k)),
\label{eqn:Psit2}
\end{eqnarray}
since $\psi_{\pm}(\hat{\vp})$ are eigenfunctions
of ${\cal H}(\vp)$ with the eigenvalues
$\lambda=\pm p$.
It should be noted that the time dependence and 
the initial-qubit dependence are separately controlled
by the absolute value of the vector $\vq(k)$,
$q(k)=|\vq(k)|$, and by its direction,
$\hat{\vq}(k)$, respectively.

\subsection{Weak limit theorem}

We recall the formula (\ref{eqn:expec2}).
It is enough to consider the case
$
  f(x)=x^r, r=0,1,2, \cdots.
$
\begin{widetext}
Following the argument by Grimmett {\it et al.} \cite{GJS04},
we find from (\ref{eqn:Psit2}),
\begin{eqnarray}
\left(i \frac{d}{dk} \right)^{r} \hat{\Psi}_{n}(k)
&=& \left(\frac{d q(k)}{dk} \right)^{r} 
e^{-i q(k) n} C_{+}(\hat{\vq}(k)) \psi_{+}(\hat{\vq}(k)) 
n^{r} \nonumber\\
&+& \left( - \frac{d q(k)}{dk} \right)^{r} 
e^{i q(k) n} C_{-}(\hat{\vq}(k)) \psi_{-}(\hat{\vq}(k)) n^{r}
+{\cal O}(n^{r-1}). 
\label{eqn:diff1}
\end{eqnarray}
Insert (\ref{eqn:Psit2}) and (\ref{eqn:diff1})
in the formula (\ref{eqn:expec2}),
use the orthonormality (\ref{eqn:eigenf3}), and
take the limit $n \to \infty$. Then we have
\cite{GJS04},
\begin{equation}
\lim_{n \to \infty} \langle (X_{n}/n)^{r} \rangle
\nonumber\\
= \int_{-\pi}^{\pi} \frac{dk}{2 \pi} \Bigg\{
|C_{+}(\hat{\vq}(k))|^2
+(-1)^{r}|C_{-}(\hat{\vq}(k))|^2 \Bigg\}
\left(
\frac{\sin (k+\theta)}
{\sqrt{1/u^2-\cos^2(k+\theta)}}
 \right)^{r},
\label{eqn:limit1}
\end{equation}
where we used (\ref{eqn:dqdk}).

Now we perform the map $k \to \gamma$.
By (\ref{eqn:sing}), (\ref{eqn:dkdg}) and 
(\ref{eqn:Cpmq}), we have
\begin{eqnarray}
\lim_{n \to \infty}
\langle (X_{n}/n)^{2m} \rangle
&=&
\int_{-\pi}^{\pi} \frac{d \gamma}{2 \pi}
\frac{\sqrt{1-u^2}}{1-u^2 \sin^2 \gamma}
(u \sin \gamma)^{2m}, \nonumber\\
\lim_{n \to \infty}
\langle (X_{n}/n)^{2m+1} \rangle
&=&  - \left\{ (|\alpha|^2-|\beta|^2)
+ \frac{\sqrt{1-u^2}}{u}
(\alpha \beta^{*} e^{-i(\phi-\theta)}
+ \alpha^{*} \beta e^{i(\phi-\theta)}) \right\}
\nonumber\\
&& \qquad \times
\int_{-\pi}^{\pi} \frac{d \gamma}{2 \pi}
\frac{\sqrt{1-u^2}}{1-u^2 \sin^2 \gamma}
(u \sin \gamma)^{2m+2}
\nonumber
\end{eqnarray}
for $m=0,1,2, \cdots$.
Then we change the variable as (\ref{eqn:defy}).
The results are expressed using
(\ref{eqn:Konno3}) as
\begin{eqnarray}
\lim_{n \to \infty}
\langle (X_{n}/n)^{2m} \rangle
&=& \int_{-u}^{u} dy \,
\mu(y; u) y^{2m}, \nonumber\\
\lim_{n \to \infty}
\langle (X_{n}/n)^{2m+1} \rangle
&=&  - \left\{ (|\alpha|^2-|\beta|^2)
+ \frac{\sqrt{1-u^2}}{u}
(\alpha \beta^{*} e^{-i(\phi-\theta)}
+ \alpha^{*} \beta e^{i(\phi-\theta)}) \right\}
\int_{-u}^{u} dy \,
\mu(y; u) y^{2m+2}.
\label{eqn:limit3}
\end{eqnarray}
\end{widetext}
Then we have arrived at the following limit theorem,
which was first obtained by Konno \cite{Konno02a,Konno02b}.

\vskip 0.3cm
\noindent {\bf Theorem} \quad
Let $X_{n}$ be the one-dimensional quantum walk
associated with the matrix
\begin{eqnarray}
A &=& \left( \matrix{ a & b \cr -b^{*} & a^{*} } \right)
\nonumber
\end{eqnarray}
with $a,b \in \C, |a|^2+|b|^2=1$, 
which starts from the state
$$
\Psi_{0}(x)=\delta(x) 
\left( \matrix{ \alpha \cr \beta } \right),
\quad
|\alpha|^2+|\beta|^2=1, \quad \alpha, \beta \in \C.
$$
Then for any analytic function $f(x)$ on $\Z$,
$$
\langle f(X_{n}/n) \rangle
\quad \to \quad
\int_{-\infty}^{\infty} dy \, f(y) \nu(y)
\quad \mbox{in} \quad n \to \infty,
$$
where $\nu(y)$ is given by (\ref{eqn:Konno2})-(\ref{eqn:Konno4}).

\section{CONCLUDING REMARKS}

We remark that the quantity 
$h=\vsigma \cdot \hat{\vp}$ is called the 
{\it helicity}. The energy-momentum eigenfunctions
$\psi_{+}(\vp)$ and $\psi_{-}(\vp)$ are also
the eigenfunctions of the helicity with
the eigenvalues $+1$ and $-1$, respectively.
The difference between the behaviors of even moments
and odd moments in (\ref{eqn:limit3}) comes from the
fact that these two states with opposite helicity
contribute with the same sign to even moments
and with alternating signs to odd moments,
respectively, as we can see in (\ref{eqn:limit1}).
In the quantum spin systems, the term of the form
$\vsigma \cdot {\bf H}$ and $\vsigma \cdot {\bf d}$
appear in Hamiltonian, representing the
Zeeman interaction between a spin and an external magnetic field
${\bf H}$, and the spin-orbit interaction with
an anisotropy vector ${\bf d}$, respectively.
We expect that
these analogue with other physical systems
will be useful, when we consider many-body systems
of quantum walkers in the future.

The present one-dimensional quantum walks 
make a family of models,
each of which is specified by a $2 \times 2$ unitary matrix.
The corresponding family of orbital states 
in the three-dimensional
momentum space studied in this paper can be regarded as
a geometrical representation of the unitary
group ${\rm U}(2)$. This point of view 
will provide a useful sight
in quantum-walk problems not only in solving problems
but also in setting problems themselves 
and in classifying results.
For example, what kind of quantum random walk models
can be associated with the unitary group ${\rm SU}(3)$,
in which the so-called Gell-Mann's eight matrices
(see, for example, \cite{Geo99})
play the similar role to the Pauli matrices
in ${\rm SU}(2)$ ?
In the recent papers, a lot of interesting
phenomena have been reported in a variety
of quantum random walk models, which are associated with
$3 \times 3$ or larger matrices \cite{IKK04,IK05,VBBB05}.
Symmetry and invariance, which are sometimes
hidden in a variety of phenomena 
observed in quantum walks,
should be clarified using the group-theoretical
investigation in the future.

\appendix

\begin{widetext}
\section{ON THE ORBITS}

Here we assume that
$k + \theta \in [-\pi, \pi)$
for simplicity of description.
By (\ref{eqn:q01}), we can see that
\begin{eqnarray}
&(i)& 
q(k)=\pi - \arccos u \searrow \frac{\pi}{2} \searrow
\arccos u 
\quad \mbox{as} \quad
k+\theta=-\pi \nearrow - \frac{\pi}{2} \nearrow 0,
\nonumber\\
&(ii)& 
q(k)=\arccos u \nearrow \frac{\pi}{2} 
\nearrow \pi - \arccos u 
\quad \mbox{as} \quad
k+\theta=0 \nearrow \frac{\pi}{2} \nearrow \pi,
\nonumber
\end{eqnarray}
where $x=a \nearrow b$ 
(resp. $x=a \searrow b$) means 
the variable $x$ monotonically increases
(resp. decreases) from $a$ to $b$.
That is,
$dq(k)/dk \leq 0$, when
$-\pi \leq k+ \theta \leq 0$, and
$dq(k)/dk \geq 0$, when 
$0 \leq k+\theta < - \pi$.
By this observation and the general formula
$(d/dx) \arccos x= \mp 1/\sqrt{1-x^2}$,
we can determine the derivative as
\begin{equation}
\frac{dq(k)}{dk}=
\frac{\sin (k+\theta)}
{\sqrt{1/u^2-\cos^2(k+\theta)}}.
\label{eqn:dqdk}
\end{equation}
The equations (\ref{eqn:q01})-(\ref{eqn:q03}) give
\begin{eqnarray}
\vq(-\pi-\theta) &=& 
q_{\rm max} \Big( \sin(\phi-\theta), \cos(\phi-\theta), 0 \Big), 
\nonumber\\
\vq(-\pi/2-\theta) &=&
\frac{\pi}{2}
\Big( \sqrt{1-u^2} \cos (\phi-\theta), 
- \sqrt{1-u^2} \sin (\phi-\theta), u \Big),
\nonumber\\
\vq(-\theta) &=& 
-q_{\rm min} \Big( \sin(\phi-\theta), \cos(\phi-\theta), 0 \Big),
\nonumber\\
\vq(\pi/2-\theta) &=&
\frac{\pi}{2}
\Big( -\sqrt{1-u^2} \cos (\phi-\theta), 
\sqrt{1-u^2} \sin (\phi-\theta), -u \Big),
\nonumber
\end{eqnarray}
where 
$q_{\rm min}= \arccos u$ and 
$q_{\rm max}=\pi - q_{\rm min}$.
\end{widetext}

The relations (\ref{eqn:cosg}) and (\ref{eqn:sing}) imply
\begin{eqnarray}
k+\theta = -\pi \quad
&\Leftrightarrow& \quad \gamma=-\pi, \nonumber\\
k+\theta = -\pi/2 \quad
&\Leftrightarrow& \quad \gamma=\pi/2, \nonumber\\
k+\theta = 0 \quad
&\Leftrightarrow& \quad \gamma=0, \nonumber\\
k+\theta = \pi/2 \quad
&\Leftrightarrow& \quad \gamma=-\pi/2, \nonumber
\end{eqnarray}
and they can be inverted as
\begin{eqnarray}
\label{eqn:sinkt}
\sin(k+\theta) &=& - \frac{(\sqrt{1-u^2}/u) \sin \gamma}
{\sqrt{(1-u^2)/u^2+ \cos^2 \gamma}}, \\
\label{eqn:coskt}
\cos(k+\theta) &=& 
 \frac{ (1/u) \cos \gamma}
{\sqrt{(1-u^2)/u^2+ \cos^2 \gamma}}.
\end{eqnarray}
By differentiate the both sides of (\ref{eqn:coskt}),
we have
$$
-\sin (k+\theta) dk =
\frac{d}{d \gamma} \left[
\frac{(1/u) \cos \gamma}
{\sqrt{(1-u^2)/u^2+ \cos^2 \gamma}} \right] d \gamma.
$$
Here we see
$$
\frac{d}{d \gamma} \left[
\frac{(1/u) \cos \gamma}
{\sqrt{(1-u^2)/u^2+ \cos^2 \gamma}} \right] 
= \sin (k+\theta)
\frac{\sqrt{1-u^2}}{1-u^2 \sin^2 \gamma},
$$
where we have used (\ref{eqn:sinkt}).
Then the Jacobian $J=|dk/d \gamma|$
is determined as (\ref{eqn:J}).

\section{INITIAL STATE DEPENDENCE}

Here we consider a two-component unit vector
$$
\phi_{0}=\left( \matrix{\alpha \cr \beta} \right),
\qquad \alpha, \beta \in \C,
\quad
|\alpha|^2+|\beta|^2=1.
$$
This can be represented as a linear combination of
the two eigenfunctions of 
${\cal H}(\vp)=\vsigma \cdot \vp$ 
given by (\ref{eqn:eigenf2}) as 
$$
\phi_{0}=C_{+}(\hat{\vp}) \psi_{+}(\hat{\vp})
+C_{-}(\hat{\vp}) \psi_{-}(\hat{\vp}),
$$
\begin{widetext}
where
\begin{eqnarray}
C_{+}(\hat{\vp}) &=& \psi_{+}^{\dagger}(\hat{\vp}) \phi_{0}
= \alpha \cos \frac{\theta_{p}}{2}
+ \beta \sin \frac{\theta_{p}}{2} e^{-i \varphi_{p}},
\nonumber\\
C_{-}(\hat{\vp}) &=& \psi_{-}^{\dagger}(\hat{\vp}) \phi_{0}
= - \alpha \sin \frac{\theta_{p}}{2} e^{i \varphi_{p}}
+ \beta \cos \frac{\theta_{p}}{2}
\label{eqn:Cpmp1}
\end{eqnarray}
by the orthonormality (\ref{eqn:eigenf3}).
By straightforward calculation we find
$$
|C_{\pm}(\hat{\vp})|^2 
= \frac{1}{2}
\pm \frac{1}{2} \Bigg\{
(|\alpha|^2-|\beta|^2) \hat{p}_{3}
+\alpha \beta^{*}(\hat{p}_{1}+i \hat{p}_{2}) 
+\alpha^{*} \beta (\hat{p}_{1}-i \hat{p}_{2}) \Bigg\}.
$$

Now we set $\hat{\vp} \Rightarrow \hat{\vq}(\gamma)$, which is
given by (\ref{eqn:q3qg}).
Since 
$$
\hat{p}_{1} \pm i \hat{p}_{2} \quad \Rightarrow \quad
\hat{q}_{1} \pm i \hat{q}_{2} 
= \Big( \sqrt{1-u^2} \sin \gamma \mp 
i \cos \gamma \Big) e^{\mp i (\phi-\theta)},
$$
the above results become
\begin{eqnarray}
|C_{\pm}(\hat{\vq}(\gamma))|^2 
&=& \frac{1}{2} \pm \frac{1}{2}
\Big\{ (|\alpha|^2-|\beta|^2) +
\frac{\sqrt{1-u^2}}{u}
(\alpha \beta^{*} e^{-i(\phi-\theta)}
+ \alpha^{*} \beta e^{i(\phi-\theta)} \Big\} u \sin \gamma
\nonumber\\
&& \mp \frac{1}{2} i
(\alpha \beta^{*} e^{-i(\phi-\theta)}
- \alpha^{*} \beta e^{i (\phi-\theta)} )
\cos \gamma.
\label{eqn:Cpmq}
\end{eqnarray}

\begin{acknowledgments}
The present authors thank H. Tanemura,
T. Sasamoto, and T. Oka for useful discussions
on quantum random walk models and related problems.
\end{acknowledgments}

\end{widetext}


\end{document}